# Spin-phonon interaction increased by compressive strain in antiferromagnetic MnO thin films

Alireza Kashir[1], Veronica Goian[2], Oliva Pacherová[2], Yoon Hee Jeong[1,3], Gil-Ho Lee[1*], Stanislav Kamba[2*]

[1]Department of Physics, Pohang University of Science and Technology (POSTECH), Pohang, 37673, Republic of Korea
[2]Institute of Physics, Czech Academy of Sciences, Na Slovance 2, 182 21 Prague 8, Czech Republic
[3]Department of Physics, Korea Advanced Institute of Science and Technology, Daejon, 34141, Republic of Korea

**Abstract**
MnO thin films with various thicknesses and strains were grown on MgO substrates by pulsed laser deposition, then characterized using x-ray diffraction and infrared reflectance spectroscopy. Films grown on (001)-oriented MgO substrates exhibit homogenous biaxial compressive strain which increases as the film thickness is reduced. For that reason, the frequency of doubly-degenerate phonon increases with the strain, and splits below Néel temperature $T_N$ due to the magnetic-exchange interaction. Films grown on (110)-oriented MgO substrates exhibit a huge phonon splitting already at room temperature due to the anisotropic in-plane compressive strain. Below $T_N$, additional phonon is activated in the IR spectra; this trend is evidence for a spin-order-induced structural phase transition from tetragonal to monoclinic phase. Total phonon splitting is 55 cm$^{-1}$ in (110)-oriented MnO film, which is more than twice the value in bulk MnO. This result is evidence that the nearest neighbor exchange interaction, which is responsible for the magnetically driven phonon splitting, is greatly increased in compressively strained films.

*Corresponding Authors
kamba@fzu.cz
lghman@postech.ac.kr

**Introduction**

Magnetic ordering can affect the lattice dynamics of strongly-correlated electronic systems This phenomenon can yield phonon anomalies or even splitting of phonons below the Néel temperature $T_N$ in various antiferromagnets [1,2,3,4]. Phonon anomalies have also been observed near $T_N$ in most multiferroics (regardless of whether they are type I or type II) [5,6,7]. Coupling of polar phonons with spin waves is also responsible for activation of electrically active magnons (called electromagnons) in the terahertz dielectric spectra of multiferroics [8,9].

A huge degree (up to 10%) of phonon splitting can occur in magnetic binary oxides at the onset of magnetic alignment [1,2,3]. It motivated the researchers to investigate the origin of this phenomenon in more details. The magnetic-order-induced phonon splitting in manganese monoxide MnO was first predicted theoretically and *ab initio* calculation showed a substantial spin-order-induced anisotropy for the Born effective charge tensor and thus for the zone-center optic phonons [10]. This work suggested the phenomenon of *solely magnetically-induced phonon splitting*; several subsequent theoretical studies attempted to explain the nature of spin-phonon coupling in detail. Anisotropic magnetic coupling due to the non-cubic distribution of the spin density can produce a splitting of transversal optical phonons during the superexchange interaction in MnO and NiO [11]. The degree of phonon splitting increases linearly with increase in the non-dominant magnetic exchange interaction $J_1$ in transition-metal monoxides and in frustrated antiferromagnetic (AFM) oxide spinels [3].

Phonon splitting must be a consequence of a decrease in crystal symmetry below $T_N$. At room temperature (RT), MnO crystalizes in the cubic rocksalt structure (space group $Fm\bar{3}m$), with lattice parameter of 4.445 Å [12]. At $T_N$ = 118 K, MnO transforms from paramagnetic to AFM phase. In the AFM structure the magnetic moments are arrayed in ferromagnetic sheets parallel to (111) planes, and the direction of magnetization in neighboring planes is antiparallel [13]. The low-temperature crystal structure of MnO is a matter of debate; it seems to undergo spin-order-induced symmetry lowering to rhombohedral structure [14,15,13,12], but structural investigations of AFM phase using neutron diffraction revealed a monoclinic crystal structure with the acentric space group $C2$ [16] or centrosymmetric $C2/m$ [17]. Even the high-temperature cubic structure of MnO has been questioned, because new high-resolution synchrotron radiation studies revealed tetragonal structure with the space group $I4/mmm$ [18]. This structure is rather plausible, because it has group-subgroup relation with the monoclinic structure that is observed below $T_N$.

In high-temperature cubic structure, only one triple-degenerated optical phonon with symmetry $F_{1u}$ should be infrared (IR) active [2]. The symmetry permits no Raman or silent mode. If the paraelectric structure is tetragonal [18], the $F_{1u}$ phonon should split in two components $A_{2u}$ and $E_u$, and both modes should be also IR active. Room-temperature IR reflectivity spectra of MnO bulk samples revealed two reflection bands; the high-frequency mode near 410 cm$^{-1}$ was explained by a multiphonon absorption,[19,20,2] so only the peak near 270 cm$^{-1}$ indicates an optical phonon in the paramagnetic phase; this observation supports the hypothesis of cubic symmetry. Nevertheless, finite phonon damping occurs, so one cannot exclude a small phonon splitting due to small ($10^{-4}$)[18] tetragonal distortion, which cannot be revealed in the IR spectra. Inelastic neutron scattering revealed 10% phonon splitting in doublet below $T_N$ [21], and IR reflectivity even found splitting into three components; [1,2,3] these observations support the hypothesis of monoclinic symmetry.

A theoretical study of Wan *et al.* [22] has predicted that 5% compressive biaxial strain can increase magnetic exchange interaction, and thereby induce ferroelectric order in the AFM phase of thin MnO film. This study proposed a completely new mechanism for preparation of strain-induced multiferroics. Epitaxial strain has been used to induce ferroelectricity in quantum paraelectric SrTiO$_3$ [23] or in antiferromagnets with large spin-phonon coupling (EuTiO$_3$ and SrMnO$_3$) [6,24], but in all of these materials the ferroelectric phase transition is of displacive type, i.e., driven by a polar soft phonon. Moreover, in strained EuTiO$_3$, the AFM order changes to ferromagnetic order under strain, so high magnetoelectric coupling is expected in this type-I multiferroic material. In type-II multiferroics, the ferroelectric polarization can be induced by a special spin order, as has been observed in many magnetic materials (e.g. review [25]), but spin-order induced multiferroicity under epitaxial strain has been predicted only in MnO [22]. Using an *ab initio* study, Fischer *et al.*[26] calculated the effect of compressive strain on the nearest neighbor magnetic interaction in transition metal monoxides. They showed that $J_1$ in MnO substantially increases as a function of compressive strain. Considering the experimental observation of Kant *et. al,*[3] one expects to see a substantial increase of spin-phonon interaction under compressive strain, because there is a linear dependence between the degree of splitting and non-dominant exchange coupling constant $J_1$.

The aim of this article is to study the influence of the strain and AFM ordering on lattice dynamics in MnO thin films. The phonons are IR active and IR reflectance spectroscopy is highly sensitive to polar phonons in ultrathin films,[27] so we use this technique between 10 and 300 K.

**Experimental details**

To apply different level of strain in various directions in MnO, MgO single crystals (Crystech) oriented in 001 and 110 directions were selected as the substrates in this research. All substrates were cleaned ultrasonically in acetone and methanol for 10 minutes in each to remove the contaminations from the top surface. After drying under a high purity nitrogen gas flow, the annealing process was carried out according to the Table 1, to prepare atomically smoothed surface [28].

Table 1. Annealing conditions for MgO substrates with different orientations.

| Substrate | Atmosphere | Temperature | Time |
|---|---|---|---|
| **MgO (001)** | air | 1150 °C | 3 hr |
| **MgO (110)** | air | 950 °C | 3 hr |

A MnO ceramic pellet (99.99%) was used as a target for pulsed laser deposition (PLD) of all thin films. A KrF pulsed laser ($\lambda$ = 248 nm) with a repetition rate of 10 Hz was operated to ablate the ceramic pellet. During the deposition process, the substrate was heated to 650 °C, which is the optimal temperature to grow MnO films of highest quality [28]. The MnO films were grown under vacuum of $10^{-6}$ Torr. The laser beam was focused on the target surface through an optical lens at a fluence $1 \leq F \leq 2$ J/cm$^2$. The substrate was placed 50 mm above the target. MnO films with thicknesses from 17 nm to 65 nm were deposited, then at the end of deposition process, the chamber was immediately cooled to RT at 10 °C/min under the growth condition.

The phases of the films and their crystal quality were evaluated using an X-ray diffractometer with a copper source operating at 40 V and 200 mA, followed by a theta rocking scan around the detected MnO Bragg peaks. An X-ray reflectometry technique was used to determine the film thickness. Moreover, X-ray diffraction studies to determine film thicknesses, strain and lattice parameters of MnO grown on MgO (110) substrate were performed using a Bruker D8 Discover equipped with rotating Cu anode ($\lambda$(CuK$\alpha$1) = 1.540598 Å; $\lambda$(CuK$\alpha$2) = 1.544426 Å) working with 12-kW power. A parabolic Göbel mirror was located on the side of the incident beam. Analyzer Soller slits, analyzer crystal (200-LiF) or dual analyzer "Pathfinder" were used on the side of the diffracted beam. The out-of-plane lattice parameter $d_{110}$ was obtained using a symmetrical 2θ/θ scan; $d_{100}$ and $d_{010}$ lattice parameters were obtained using inclined ($\chi$= 45 °) symmetrical 2θ/θ scans. The $d_{001}$ lattice parameter was found using a reciprocal space map (RSM) of the 224 diffraction.

The IR reflectivity measurements were performed using a Bruker IFS 113v Fourier transform IR spectrometer equipped with a helium-cooled (1.6 K) silicon bolometer. Spectra were measured at temperatures from 10 to 300 K, which were maintained using an Oxford Optistat CF cryostat with 3-mm-thick polyethylene windows transparent up to 650 cm$^{-1}$. RT spectra were taken up to 4000 cm$^{-1}$. Polarized and unpolarized reflectance measurements were performed in near-normal incidence geometry. Metal mesh deposited on 6-μm-thick mylar was used as a polarizer.

**Results and discussion**

### A. X-Ray diffraction analysis

Out-of plane lattice constant, strain, thickness and sample homogeneity of all thin films were determined using XRD at 300 K. In-plane strain and lattice parameters of two MnO(110) films with thicknesses 26.5 and 18 nm were also measured.

Fluence of the laser affected the XRD patterns of films (Fig. 1a). As fluence was increased to 2 J/cm$^2$, a second phase ($Mn_3O_4$) emerged, possibly because the increased energy effected the ionization of Mn atoms. Pure MnO phase was obtained by reducing fluence below 1.5 J/cm$^2$. As film thickness was decreased, the 2θ angle of the MnO 002 peak decreased (Fig. 1b); this change implies an out-of plane expansion of the crystal structure of MnO. This expansion might be a result of the lattice mismatch between MnO and MgO. The theoretical lattice mismatch between film and substrate is relatively high (about 5%); the critical thickness for total relaxation decreases as theoretical lattice mismatch increases [29]. Growth in vacuum also causes the relaxation of film as there is no decelerating force to reduce the energy of ablated species which, in turn, provides driving force for the surface diffusion.

In 26.5-nm-thick MnO/MgO (110), the two RSMs around the peak positions in the 224 diffraction (Fig. 2a) had very different intensities of substrate and layer peaks, so the RSMs were measured independently, then combined into one image that has a common scale. The double peak at 224 coordinates of MgO substrate is due to the CuK$\alpha_{12}$ doublet of the incident radiation. After a processing of the layer peak, the coordinates of the focal point $L_{lay}$ and $H_{lay}$ were obtained and the corresponding values of $d_{110}$ and $d_{001}$ could be derived. The value of $d_{110}$ can be compared with its value that was obtained by direct measurement using a symmetrical 2θ/θ scan. Fig. 2b-d) show diffractions 220, 200 and 020, and X-ray reflectivity of 26.5 nm thick MnO(110) film. These scans lead directly to the $d_{110}$, $d_{100}$ and $d_{010}$ lattice parameters and to the layer thickness. The same procedure was used for two MnO(110) films (thickness $18 \pm 1$ nm and $26.5 \pm 0.5$ nm). The features of the scans indicate that the crystallinity of the film increases as its thickness increases. This inference is confirmed by the high uncertainty of the 18-nm film.

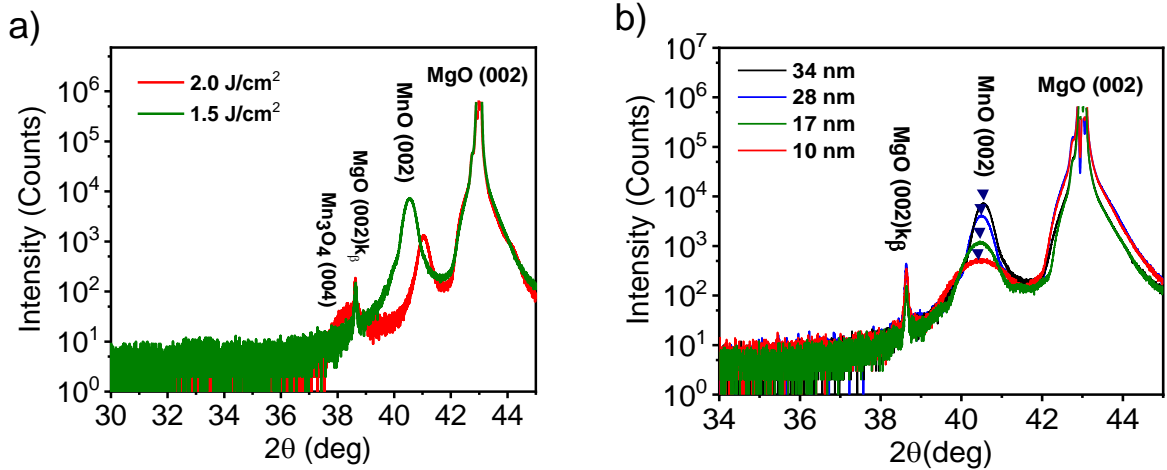

Fig. 1 (a) XRD spectra of the films deposited on MgO (001) substrate using different energy fluences. The film grown with laser fluence of 2.0 J/cm² contains $Mn_3O_4$ inclusions. b) XRD spectra of the films grown on MgO (001) with different thicknesses.

All lattice parameters in two MnO(110) films were determined from the RSM: 26.5-nm-thick film had $c = 4.409 \pm 0.003$ Å and $a = b = 4.455 \pm 0.001$ nm; 18-nm-thick MnO (110) film had $c = 4.390 \pm 0.005$ Å and $a = b = 4.460 \pm 0.002$ Å. Bulk MnO has $a = 4.445$ Å [12], so in both films the $c$ parameter is smaller than in the MnO bulk (compressive strain) and $a$, $b$ parameters are larger in the thinnest 18-nm-thick film than in the bulk MnO.

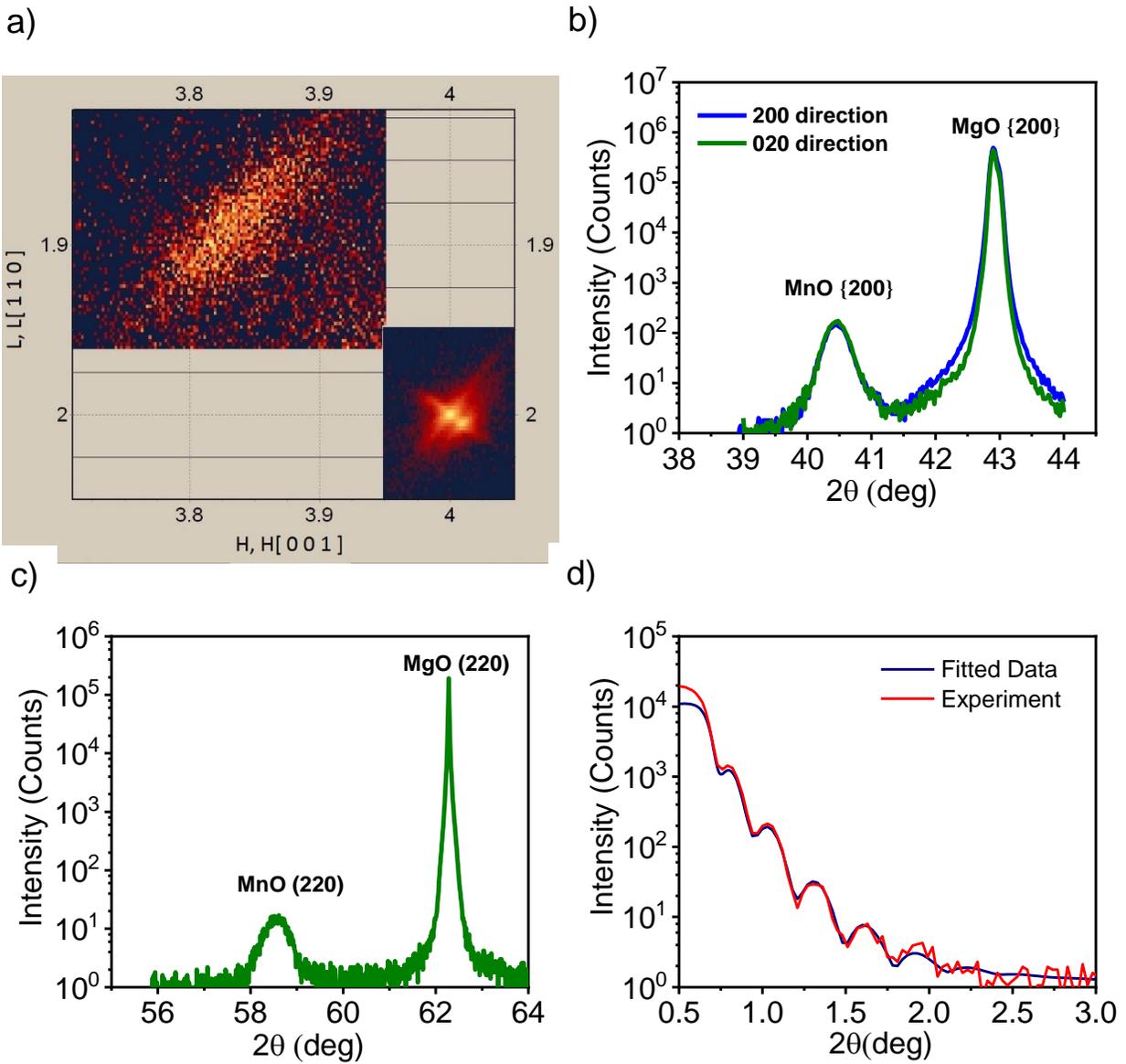

Fig. 2. (a) X-ray reciprocal space map (RSM) of 26.5 nm thick MnO/MgO(110) around 224 diffraction. Double peak seen at 224 is caused by CuKα$_{12}$ doublet of Cu rotating anode. (b,c,d) X-ray scans of the diffractions 200, 020 and 220, and of the reflectivity for the sample with the film thickness 26.5 nm.

### B. IR reflectivity spectra

To determine how the strain affects lattice dynamics, films with various thicknesses grown on MgO substrates with (001) and (110) orientations were measured. Selected examples of the room-temperature reflectance spectra are shown in Fig. 3. Besides strong phonon response of MgO substrate (frequencies of MgO excitations are marked by arrow), reflection peaks from MnO phonons are seen near 285 cm$^{-1}$. Phonon frequencies in the films correspond to reflectance maxima [30], whereas in bulk substrate, transverse optical phonon frequency corresponds to the low-frequency edge of the reflection band [1,19]. Multiphonon excitation frequency in the MgO substrate corresponds to the feature near 650 cm$^{-1}$ (arrow, Fig. 3a). Near-normal IR reflectivity gives only in-plane response, so we cannot see phonons polarized perpendicularly to the film plane, i.e., in the [001] direction. As a result of the epitaxial strain in the film induced by the lattice mismatch between the MgO substrate and MnO film, this phonon can have different frequency than phonons polarized in (001) plane. In-plane strain is homogeneous, so films grown on MgO(001) exhibit a single peak (Fig. 3a) at RT; this result means that phonons polarized along [100] and [010] have the same frequency (i.e., the phonon is doubly degenerate). Intensity of the MnO IR reflection peak increased as the thickness of the films was increased. Strain is relaxed in 35-nm-thick film, but is present in thinner film and increases as thickness is reduced.

IR spectra of the MgO(110) substrate (Fig 3b) are the same as in the (001) plane (Fig 3a) because MgO is cubic. Nevertheless, the spectra of MnO thin films grown on MgO(110) (Fig. 3b) are completely different than of MnO (001) films (Fig. 2a), because the (110) oriented substrate induces anisotropic in-plane strain, and the MnO phonon has already split at RT. The splitting decreases as the thickness of the films increases (Fig. 3b), because the strain decreases as film thickness increases. In the 65-nm thick film, the strain had relaxed, so the splitting disappeared.

The intensities of the reflection peaks are increased in polarized spectra (Fig. 3b, inset), because these modes are selectively excited, whereas in the unpolarized spectrum, both modes are simultaneously excited. The phonon polarized along [001] has higher frequency than the mode polarized perpendicularly to [001], because the *c* lattice parameter is smaller than *a* and *b* lattice parameters (see XRD results).

IR reflectance spectra of the film are highly sensitive to crystal structure and phase impurities in the film. One film had a completely different IR spectrum (Fig. 3c) than other films. Two reflection minima appear in the reststrahlen band of MgO near 470 and 598 cm$^{-1}$, and correspond to phonons in Mn$_3$O$_4$ [31]. The film probably contains more Mn$_3$O$_4$ phase than MnO, so the MnO reflection band near 280 cm$^{-1}$ is weak and smeared due to its high damping and low intensity. Mn$_3$O$_4$ phase impurity was also confirmed by independent XRD measurement (Fig. 1a).

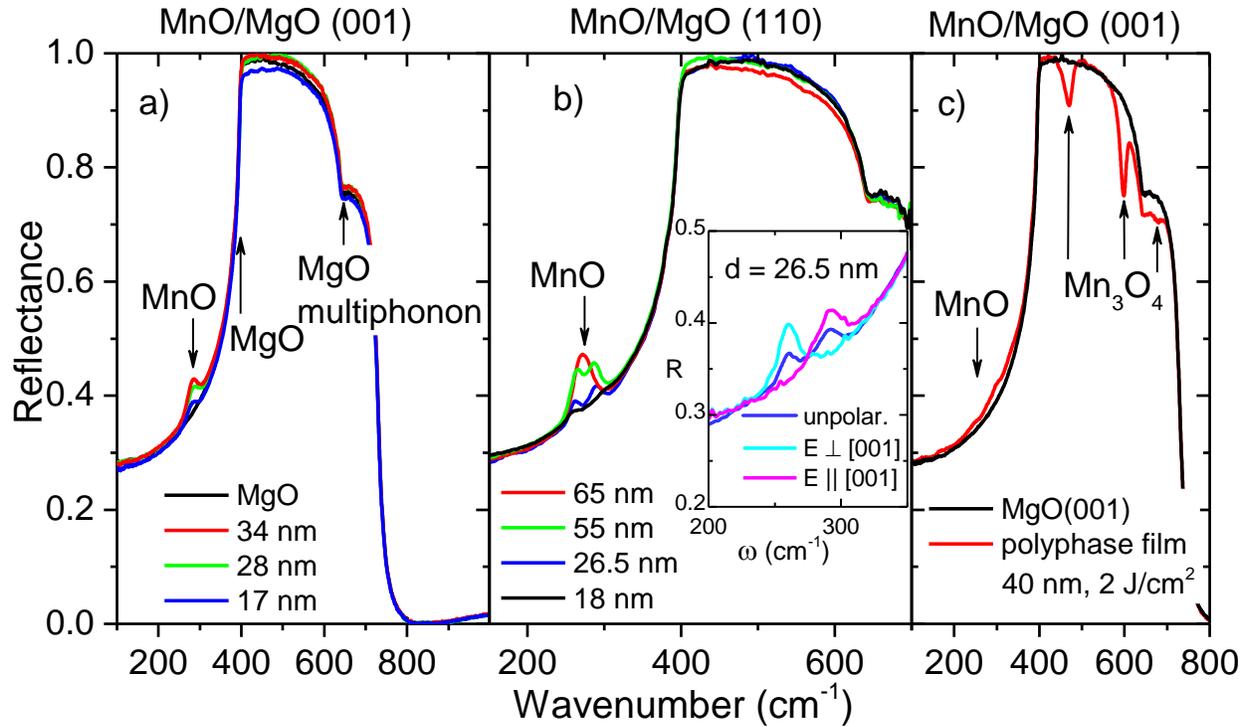

Fig.3. Unpolarized room-temperature IR reflectance spectra of MnO thin films with various thicknesses and strains grown on a) (001) and b) (110)-oriented MgO substrates. Inset: unpolarized spectrum of 26.5-nm-thick MnO (110) film compared with two IR polarized spectra. c) IR reflectance spectrum of MgO (001) substrate compared with the spectrum of poly-phase thin film, in which $Mn_3O_4$ phase dominates.

Temperature dependence of IR reflectance spectra of selected samples are shown in Fig. 4, temperature evolution of MnO phonon frequencies is plotted in Fig. 5. In MnO/MgO(001) films, an doubly-degenerate $E_u$ phonon splits into a doublet below $T_N$. In the paramagnetic phase, MnO phonon frequency increases with the biaxial compressive strain in the film due to increase in spring constants between atoms. In AFM phase, the highest phonon splitting (~30 cm$^{-1}$) was observed in the thinnest film (thickness = 17 nm), because its compressive strain was highest; this splitting was 20% higher than in the bulk MnO in which splitting is 25 cm$^{-1}$ [1,3]. This result is evidence that biaxial compressive strain significantly strengthens the magnetic exchange interaction.

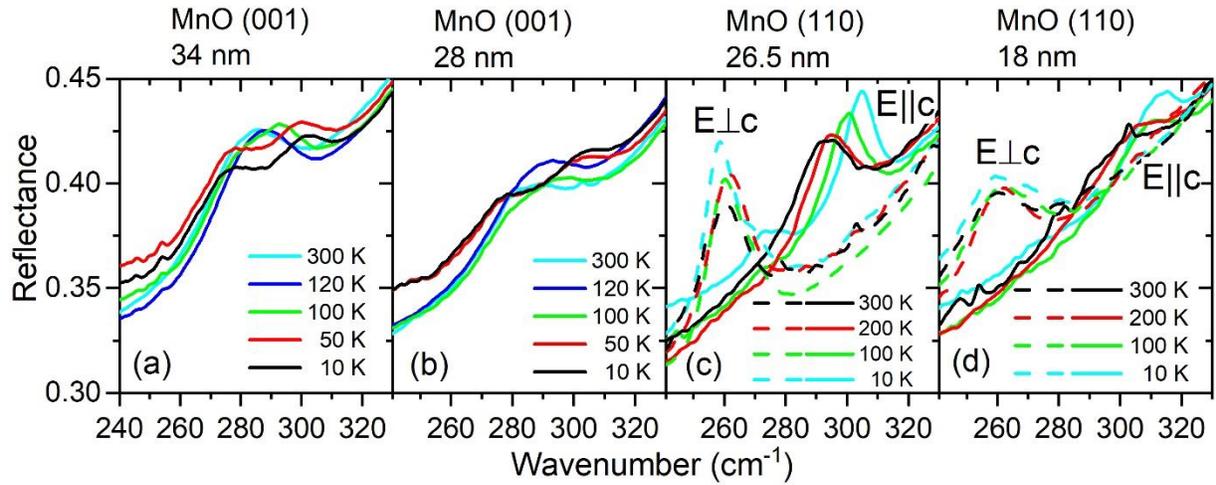

Fig. 4. Temperature dependence of IR reflectance spectra in the range of MnO phonons. (a,b) unpolarized spectra of in-plane isotropically-strained MnO (001) films on MgO (001). (c,d) IR polarized spectra of MnO (110) films grown on MgO (110). Orientations and thicknesses of the films are marked above the figures.

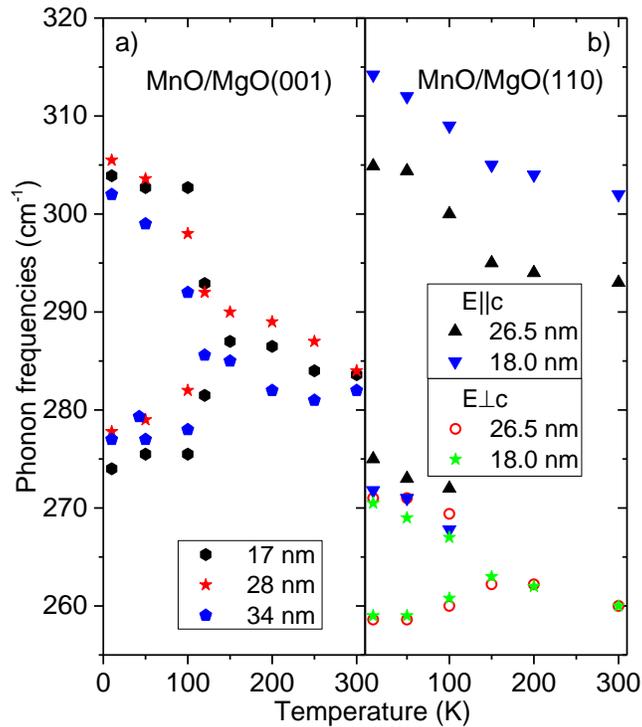

Fig. 5. Temperature dependence of polar phonon frequencies in (a) MnO (001) and (b) MnO (110) films with various thicknesses (strain) obtained from the spectra in Fig. 4. Strain is anisotropic in the (110) oriented films, so polarizations of the spectra are marked in b).

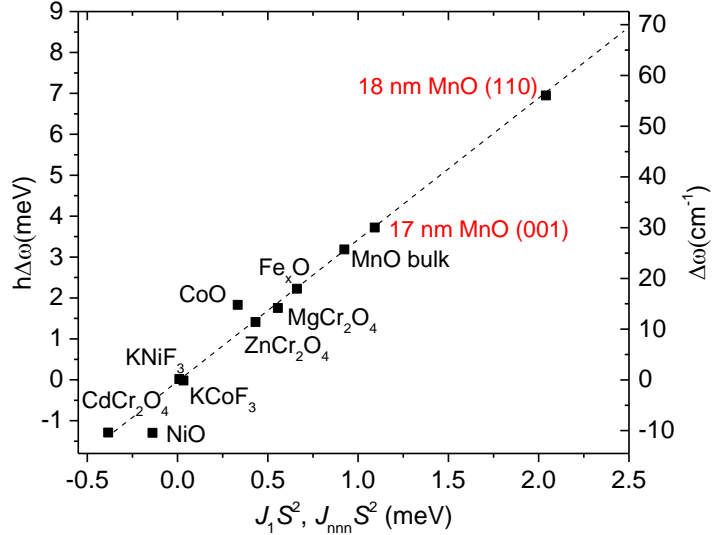

Fig. 6. Dependence of phonon splitting on non-dominant exchange contribution $J_1 S^2$ and $J_{nnn} S^2$ for various transition metal oxides and Cr spinels, respectively. The figure is adapted from Ref. 1, our observed phonon splitting in variously strained MnO thin films is marked in red.

The phonon behavior in MnO/MgO(110) is qualitatively different than in MnO(001). Due to anisotropic biaxial strain in the film, the phonon has already split in the paramagnetic phase at RT. The splitting is huge: 34 cm$^{-1}$ in 26.6-nm-thick film and 42 cm$^{-1}$ in 18-nm-thick film. The splitting increases as film thickness decreases because the reduction in thickness increases its strain-induced in-plane anisotropy. Thermal contraction causes the phonon polarized in [001] to harden slightly on cooling towards $T_N$, but exchange coupling occurs below $T_N$ so the frequency increases greatly at lower temperatures. The phonons polarized perpendicularly to [001] split below $T_N$ and the splitting again increases with increase in strain (and decrease in film thickness), to 14 cm$^{-1}$ in 18-nm-thick film. The splitting is smaller than in the bulk MnO [2], because $a$ and $b$ lattice parameters are larger in MnO(110) film than in the bulk (Section A. XRD), so the exchange coupling is reduced in $ab$ plane as Fischer *et al.* [26] predicted. Total splitting between the highest-frequency and lowest-frequency phonons extends up to 56 cm$^{-1}$, which is 124% more than in the bulk MnO. Kant *et al.* [3] discovered linear dependence between value of phonon splitting and non-dominant exchange coupling constant in various transition metal oxides and Cr spinels·; Extrapolation of that linear trend to our observed phonon splitting (Fig. 6) yields the exchange-coupling constant in strained MnO films, which is more than twice as high as that in MnO bulk.

At RT, the two phonons observed in MnO (110) films can be explained by tetragonal symmetry, which is in agreement with our XRD data. Three phonons observed at low temperatures give evidence for lower symmetry, but our IR spectra cannot be used to distinguish whether the AFM phase is orthorhombic, monoclinic or even trigonal; additional structural studies are required. Also our data cannot reveal whether the AFM phase is polar (ferroelectric) or nonpolar. This determination requires further dielectric, Raman and second-harmonic studies, which are in progress.

Finally we can summarize that IR spectroscopy is a highly sensitive tool for study of strain and temperature influences on phonons in MnO thin films (thickness down to 17 nm) grown on MgO substrates. In addition to exchange-driven phonon splitting below $T_N$, further phonon splitting was observed in the paramagnetic phase in anisotropically strained MnO (110) films grown on MgO (110). The total phonon splitting is at 10 K more than twice as strong as in MnO bulk, so the nears neighbor exchange-coupling constant $J_1$ is also more than twice as strong which is in good agreement with the ab initio study [26]. This observation supports a theoretical prediction [11] that phonon splitting is caused by exchange coupling. Our most important observation is that epitaxial strain can strengthen exchange coupling in MnO. These films under compressive strain > 5% should be studied to test whether spin-order-induced ferroelectricity occurs, as predicted theoretically by Wan *et al.* [22].

**Acknowledgments**

This work was supported by National Research Foundation (NRF) of Korea (2015R1D1A1A02062239 and 2016R1A5A1008184) funded by the Korean Government, by the Czech Science Foundation (Project No. 18-09265S) and by Operational Programme Research, Development and Education (financed by European Structural and Investment Funds and by the Czech Ministry of Education, Youth and Sports), Project No. SOLID21 - CZ.02.1.01/0.0/0.0/16_019/0000760.